\documentclass[12pt,a4paper]{article}
\usepackage{t1enc}
\usepackage{times}
\usepackage{subfig}
\usepackage{t1enc}
\usepackage{times}
\usepackage{amssymb}
\usepackage{amsmath}
\usepackage{amsthm}
\usepackage[mathscr]{euscript}
\usepackage{graphicx}
\usepackage{float}
\usepackage{enumitem}
\usepackage{mathtools}
\usepackage{hyperref}
\usepackage[totalwidth=437pt, totalheight=664pt]{geometry}

\DeclareMathAlphabet\scr{U}{scr}{m}{n}
\SetMathAlphabet\scr{bold}{U}{scr}{b}{n}
\DeclareFontFamily{U}{scr}{\skewchar\font'177}%
\DeclareFontShape{U}{scr}{m}{n}{<-6>rsfs5<6-8>rsfs7<8->rsfs10}{}%
\DeclareFontShape{U}{scr}{b}{n}{<-6>rsfs5<6-8>rsfs7<8->rsfs10}{}%
\newtheorem{satz}{Theorem}[section]
\newtheorem{prop}[satz]{Proposition}

\newtheorem{lemma}[satz]{Lemma}
\newtheorem{cor}[satz]{Corollary}
\newtheorem{@definition}[satz]{Definition}
\newenvironment{defi}{\begin{@definition}\rm}{\end{@definition}}
\newtheorem{@bsp}[satz]{Example}
\newenvironment{bsp}{\begin{@bsp}\rm}{\end{@bsp}}
\newtheorem{@assumption}[satz]{Assumption}
\newenvironment{assumption}{\begin{@assumption}\rm}{\end{@assumption}}
\newtheorem{@convention}[satz]{Convention}

\newtheorem{@remark}[satz]{Remark}
\newenvironment{bem}{\begin{@remark}\rm}{\end{@remark}}

\newtheorem{@notation}[satz]{Notation}
\newenvironment{nota}{\begin{@notation}\rm}{\end{@notation}}

\newenvironment{prf}[1][Proof]{\begin{proof}[\textsc{#1}]}{\end{proof}}

\newcommand{\F}{\scr F}

\newcommand{\sint}{\stackrel{\mbox{\tiny$\bullet$}}{}}
\newcommand{\cadlag}{c\`adl\`ag}

\newcommand{\covar}[2]{\left\langle#1,#2\right\rangle}

\newcommand{\filtr}{\left(\F_t\right)_{t\in[0,T]}}
\newcommand{\ew}[1]{E\left(#1\right)}
\newcommand{\abs}[1]{\left|#1\right|}

\newcommand{\R}{\mathbb{R}}
\newcommand{\Rp}{\R_+}

\newcommand{\C}{\mathbb{C}}
\renewcommand{\Re}[1]{\mathrm{Re}\left(#1\right)}
\renewcommand{\Im}[1]{\mathrm{Im}\left(#1\right)}

\newcommand{\levy}{L\'evy}
\renewcommand{\theta}{\vartheta}
\renewcommand{\rho}{\varrho}

\newcommand{\dstr}{$\Delta$-strategy}
\newcommand{\dstrs}{$\Delta$-strategies}

\newcommand{\rmif}{{R-i\infty}}
\newcommand{\rpif}{{R+i\infty}}
\newcommand{\Sf}{R+i\R}

\newcommand{\ds}{\displaystyle}

\newcommand{\nim}[1]{}

\begin{document} \allowdisplaybreaks[1]
\title{On the Performance of Delta Hedging Strategies in Exponential \levy{} Models}
\author{
Stephan Denkl\footnote{
Mathematisches Seminar,
Christian-Albrechts-Universit\"at zu Kiel,
Christian-Albrechts-Platz 4,
24098 Kiel, Germany,
(e-mail: denkl@math.uni-kiel.de).}
\quad Martina Goy\footnote{
KPMG AG,
Wirtschaftspr\"ufungsgesellschaft,
Audit Financial Services,
Ganghoferstra{\ss}e 29,
80339 M\"unchen, Germany,
(e-mail: mgoy@kpmg.com).
}
\quad Jan Kallsen\footnote{
Mathematisches Seminar,
Christian-Albrechts-Universit\"at zu Kiel,
Christian-Albrechts-Platz 4,
24098 Kiel, Germany,
(e-mail: kallsen@math.uni-kiel.de).}
\\\quad Johannes Muhle-Karbe\footnote{Departement Mathematik,
ETH Z\"urich,
R\"amistrasse 101,
CH-8092 Z\"urich, Switzerland,
(e-mail: johannes.muhle-karbe@math.ethz.ch).}
\quad Arnd Pauwels\footnote{
MEAG AssetManagement GmbH,
Abteilung Risikocontrolling,
Oskar-von-Miller-Ring 18,
80333 M\"unchen, Germany,
(e-mail: apauwels@meag.com)
}
}
\date{\today}
\maketitle

\begin{abstract}
We consider the performance of non-optimal hedging strategies in exponential L\'evy models. Given that both the payoff of the contingent claim and the hedging strategy admit suitable integral representations, we use the Laplace transform approach of Hubalek et al.\ \cite{hub06} to derive semi-explicit formulas for the resulting mean squared hedging error in terms of the cumulant generating function of the underlying L\'evy process. In two numerical examples, we apply these results to compare the efficiency of the Black-Scholes hedge and the model delta to the mean-variance optimal hedge in a normal inverse Gaussian and a diffusion-extended CGMY L\'evy model.\\

\end{abstract}

\section{Introduction}\label{s:statintro}
A basic problem in Mathematical Finance is how the issuer of an option can hedge the resulting exposure by trading in the underlying. In complete markets, the risk can be offset completely by purchasing the replicating portfolio. In incomplete markets, however, additional criteria are necessary to determine reasonable hedging strategies. 

A popular approach studied intensively in the literature over the last two decades is \emph{mean-variance hedging}. Expressed in discounted terms, the idea is to minimize the \emph{mean squared hedging error} 
\begin{equation}\label{eq:qh}
\ew{\left(H - c - \int_0^T \vartheta_t dS_t\right)^2}
\end{equation}
over all in some sense admissible trading strategies $\vartheta$. Here, the random variable $H$ is the payoff of the option, $c$ is the initial endowment of the investor,\footnote{Note that we suppose here that this quantity is given exogenously. Typically, it will be chosen to match some arbitrage-free price the investor receives for selling the option. However, it can also include other initial holdings or liabilities.} and $S$ is the price process of the underlying. Since the stochastic integral $\int_0^T\vartheta_tdS_t$ represents the cumulated gains from trading $\vartheta$, \eqref{eq:qh} amounts to comparing the terminal value $c+\int_0^T\vartheta_tdS_t$ of the hedging portfolio and the option's payoff in a mean-square sense. Comprehensive overviews on the topic can be found in \cite{pham2000, schweizer2001}. For more recent publications, the reader is referred to \cite{ck07} and the references therein. In particular, semi-explicit representations of the minimal mean squared hedging error have been obtained in \cite{cerny.2007, hub06} for exponential L\'evy models and in \cite{kallsen.pauwels.09a,kallsen.vierthauer.09} for affine stochastic volatility models by making use of an integral representation of the option under consideration.

However, in practice delta hedging is still prevalent, where the hedge ratio is given by the derivative of the option price with respect to the underlying. Therefore, it seems desirable to compute also the mean-squared hedging error of such a non-optimal strategy. For exponential \levy{} models, this has been done in the unpublished thesis \cite{goy07} for continuous time under the restriction that the asset price process is a martingale and, more recently, by \cite{angelini.herzel.09} in a discrete time setup, both using the approach of \cite{hub06}. The contribution of the present study is the extension of the results of \cite{goy07} to general exponential \levy{} processes and a larger class of hedging strategies. The resulting semi-explicit formulas for the corresponding hedging errors can be found in our main result, Theorem \ref{thm:MainResult}. With these results at hand, we also compare the performance of the Black-Scholes hedge and the model delta to the mean-variance optimal hedge in several illustrative examples.

This article is organized as follows. Subsequently, we describe the setup for the price process of the underlying and for the payoff function. In Section~\ref{s:dstrs}, we introduce the class of hedging strategies to which our approach applies. These \emph{$\Delta$-strategies} (cf.\ Definition~\ref{d:deltaStr}) include in particular the Black-Scholes hedge and, more generally, delta hedges in exponential L\'evy models. In Section~\ref{sec:PerfDeltaStr}, we state the main theorem on the hedging error of $\Delta$-strategies and provide a sketch of its proof on an intuitive basis. Finally, we illustrate this result by two numerical examples in Section~\ref{s:numillustr}. In the first, we compare the performance of the Black-Scholes hedge and the variance-optimal hedge in a normal inverse Gaussian \levy{} model whose parameters are inferred from a historical time series. In the second example, we study the hedging errors of the Black-Scholes strategy, the model delta hedge, and the variance-optimal strategy in a diffusion-extended CGMY model for parameters obtained by calibration to option prices. The technical proof of the main theorem is delegated to Appendix~\ref{a:Proof}.

For stochastic background and terminology, we refer to the monograph of Jacod and Shiryaev \cite{js03}. For a semimartingale $X$, we denote by $L(X)$ the set of $X$-integrable predictable processes and write $\varphi \sint X$ for the stochastic integral of a process $\varphi \in L(X)$ with respect to $X$. By $\langle X,Y \rangle$, we denote the predictable compensator of the quadratic covariation process $[X,Y]$ of two semimartingales $X$ and $Y$, provided that $[X,Y]$ is a special semimartingale (cf.\ \cite[comment after Th\'{e}or\`{e}me~2.30]{jacod79}).

\section{Model and preliminaries}\label{s:Model}
In this section, we state our assumptions on the asset price process and the payoff. Note that we use the same setup as \cite{hub06} for mean-variance hedging.

\subsection{Asset price process}
Let $T>0$ be a fixed time horizon, and denote by $(\Omega,\F,\filtr,P)$ a filtered probability space. The discounted price process~$S$ of a non-dividend paying stock is assumed to be of the form
\begin{equation*}
 S_t = S_0 e^{X_t}, \quad S_0 \in \Rp, \quad t\in[0,T],
\end{equation*}
for a L\'evy process $X$. We demand that the stock price is square-integable,
\begin{equation} \label{secMomS}
 E(S_1^2) < \infty,
\end{equation}
which is a natural requirement when using the second moment of the hedging error as a risk criterion. The entire distribution of the L\'evy process~$X$ is already determined by the law of~$X_1$, which can be characterized in terms of the \emph{cumulant generating function} $\kappa:D\rightarrow\C$, i.e., the unique continuous function satisfying
$$E\left(e^{zX_t}\right) = e^{t\kappa(z)}$$
for $z\in D := \left\{ z \in \C : E(e^{\Re{z}X_1})  < \infty  \right\}$ and $t\in\Rp$. Note that Condition~(\ref{secMomS}) implies
\begin{equation*}
 \left\{ z\in\C : 0 \leq \Re{z} \leq 2 \right\} \subset D.
\end{equation*}
Moreover, we exclude the degenerate case of deterministic $S$ by demanding that $X_1$ has non-zero variance, i.e.,
\begin{equation*}
 \kappa(2)-2\kappa(1) \neq 0.
\end{equation*}

\subsection{Laplace transform approach}\label{s:LaplaceTransAppr}
In order to derive semi-explicit formulas in concrete models, we employ the \emph{Laplace transform approach}, which is widely used in option pricing (cf., e.g., \cite{carrmadan99, raible00}) and applied by \cite{hub06} in the context of mean-variance hedging. The key assumption is the existence of an integral representation of the payoff function in the following sense.

\begin{assumption}\label{a:intReprF}
 Let the payoff~$H$ of a contingent claim be of the form $H=f\left(S_T\right)$ for some measurable function $f:(0,\infty)\rightarrow\R$, which admits the representation
 \begin{equation}\label{e:intReprF}
     f(s) = \int_{R-i\infty}^{R+i\infty} s^z \, p(z) \, dz
 \end{equation}
 for $p:\C\rightarrow\C$ and $R\in\R$ such that $x\mapsto p(R+ix)$ is integrable and
 \begin{equation}\label{e:secMomRX}
     \ew{e^{2RX_1}} < \infty.
 \end{equation}
\end{assumption}

Note that Condition~(\ref{e:secMomRX}) implies $H\in L^2(P)$, which is again a natural assumption in view of the problem at hand.

\begin{bsp}\label{ex:Call}
 Most European options allow for an integral representation as in Assumption~\ref{a:intReprF}. For example, the payoff function $f(s)=(s-K)^+$ of a European call with strike $K>0$ can be written as
 \begin{equation*}
     f(s) = \frac{1}{2\pi i} \int_{R-i\infty}^{R+i\infty} s^z \frac{K^{1-z}}{z(z-1)} \, dz
 \end{equation*}
 for arbitrary $R>1$, cf.\ \cite[Lemma~4.1]{hub06}. In general, representations of this kind can be obtained by inverting the bilateral Laplace transform of the mapping $x\mapsto f(e^{x})$ via the \emph{Bromwich inversion formula}. More details and examples can be found in \cite[Section~4]{hub06} and \cite[Chapter~3]{raible00}.
\end{bsp}

Henceforth, we consider a fixed contingent claim~$H=f(S_T)$ satisfying Assumption~\ref{a:intReprF}.\\

In the following, we will represent several objects as integrals with respect to the weight function~$p$ from Assumption~\ref{a:intReprF}. The following terminology allows to conveniently express that such integrals are well-defined.

\begin{nota}
	A measurable function $h:(R+i\R)\rightarrow\C$ is called \emph{$p$-integrable} if $$\int_{-\infty}^{\infty}\abs{h(R+ix)}\abs{p(R+ix)}\,dx < \infty.$$ Analogously, a measurable function $h:(R+i\R)\times(R+i\R)\rightarrow\C$ is called \emph{twice $p$-integrable} if $$\int_{-\infty}^{\infty}\int_{-\infty}^{\infty}\abs{h(R+ix,R+iy)}\abs{p(R+ix)}\abs{p(R+iy)}\,dxdy < \infty.$$ Moreover, for a more convenient notation we will always write $\int_{R-i\infty}^{R+i\infty}\abs{h(z)}\abs{p(z)}|dz|$ for the integral $\int_{-\infty}^{\infty}\abs{h(R+ix)}\abs{p(R+ix)}\,dx$, and analogously for double integrals.
\end{nota}
Note that in particular, a bounded function is $p$-integrable since $z\mapsto 1$ is $p$-integrable by Assumption~\ref{a:intReprF}.

\section{\texorpdfstring{\dstrs}{Delta-strategies}}\label{s:dstrs}
We now introduce the class of strategies for which we will compute the mean squared hedging error in Section~\ref{sec:PerfDeltaStr}. Moreover, we discuss the most prominent examples.

As in \cite{angelini.herzel.09}, we focus on hedging strategies which allow for a similar integral representation as the payoff function (cf.\ Section~\ref{s:LaplaceTransAppr}).

\begin{defi}\label{d:deltaStr}
 A real-valued process $\varphi$ is called \emph{\dstr} if it is of the form
 \begin{equation*}%\label{e:intRepDstr}
     \varphi_t = \int_{R-i\infty}^{R+i\infty} \varphi(z)_t \, p(z) \, dz
 \end{equation*}
 with
 \begin{equation*}%\label{e:funcGDeltaStr}
     \varphi(z)_t = S_{t-}^{z-1} g(z,t)
 \end{equation*}
 for a measurable function $g:(\Sf)\times [0,T]\rightarrow\C$ such that
 \begin{enumerate}
   \item $t\mapsto g(z,t)$ is continuous for all $z\in\Sf$, \label{e:deltaStr0}
   %\item $t\mapsto \ds\sup_{z\in\Sf} \left|g(z,t)\right|$ is finite on $[0,T]$ and bounded on $[0,s]$ for all $s<T$, \label{e:deltaStr1}
   % \item $\overline{g(z,t)} = g(\overline{z},t)$ for all $z\in$ and all $t\in[0,T]$. \label{e:deltaStr3}
   \item $z\mapsto \ds\int_0^T \left| g(z,s) \right|^2 ds$ is $p$-integrable. \label{e:deltaStr2}
 \end{enumerate}
\end{defi}

\begin{bem}
Condition~\ref{e:deltaStr0} is required to ensure the existence of a unique solution of an ordinary differential equation (ODE) with right-hand side $g(z,\cdot)$. Condition~\ref{e:deltaStr2} ascertains that $\varphi$ is integrable with respect to $S$ and that the cumulative gains $\int_0^T \varphi_t dS_t$ of the strategy possess a second moment (cf.\ Lemma \ref{l:L(S)}). As a side remark, it also implies that any \dstr{} is admissible in the sense of \cite[Section~3]{hub06}.
\end{bem}

To motivate this definition, we now show that these $\Delta$-strategies generalize the so-called \emph{delta hedges}, which are obtained by differentiating the option price with respect to the underlying in an exponential L\'evy model. We also recall that the optimizer of the mean-variance hedging problem \eqref{eq:qh} is another special case if it is computed under a martingale measure for $S$. The most important concrete example for both is the \emph{Black-Scholes strategy}, i.e., the hedge obtained by differentiating the Black-Scholes price. This strategy also allows to achieve perfect replication in the Black-Scholes model and hence minimizes \eqref{eq:qh}. However, it is important to keep in mind that it leads to a non-trivial hedging error if applied in a different L\'evy model with jumps.

\begin{bsp} \label{ex:DeltaHedge}(\textbf{Delta hedge}) Computing the derivative of a price process with respect to the underlying in an exponential \levy{} model leads to a \dstr. To see why this holds, denote by $\widetilde{S}_t = \widetilde{S}_0e^{\widetilde{X}_t}$ an exponential \levy{} process with driver~$\widetilde{X}$ and associated cumulant generating function~$\widetilde{\kappa}$ under some martingale measure $Q$. Note that due to, e.g., model misspecification, the hedge may be derived in a model differing from the one where it is eventually applied, which is why we distinguish between the processes $S$ and $\widetilde{S}$. Using Fubini's Theorem and the independence of the increments of~$\widetilde{X}$ with respect to~$Q$, the price process of the contingent claim with payoff function~$f$ in the model~$\widetilde{S}$ is given by
\begin{align*}
         E_Q\left( f(\widetilde{S}_T) \Big| \F_t\right) = E_Q\left( \left. \int_{\rmif}^{\rpif} \widetilde{S}_T^z \,p(z)\,dz \, \right| \F_t \right) & = \int_{\rmif}^{\rpif} E_Q\left(\widetilde{S}_T^z\Big|\F_t\right) p(z)\,dz\\
              & = \int_{\rmif}^{\rpif} \widetilde{S}_t^z e^{\widetilde{\kappa}(z)(T-t)} \,p(z)\,dz.
\end{align*}
The \emph{delta hedge} of the contingent claim in this pricing model is then given by $\varphi^{\Delta}(\widetilde{S}_{t-},t)$ for
\begin{equation*}
\varphi^{\Delta}(s,t) := \frac{\partial}{\partial s} \int_{\rmif}^{\rpif} s^{z} e^{\widetilde{\kappa}(z)(T-t)} \,p(z)\,dz = \int_{\rmif}^{\rpif} s^{z-1} z e^{\widetilde{\kappa}(z)(T-t)} \,p(z)\,dz,
\end{equation*}
provided that the derivative exists and that integration and differentiation can be interchanged. This is the case if $z\mapsto z e^{\widetilde{\kappa}(z)(T-t)}$ is $p$-integrable for all $t\in[0,T)$, which is satisfied if either the the distribution of driver~$\widetilde{X}$ is sufficiently regular or the payoff function is smooth enough, cf., e.g., \cite{cont.al.05}. Using the resulting hedge in the model~$S$ yields the strategy
\begin{equation}\label{eq:FormulaDeltaHedge}
\varphi^{\Delta}(S_{t-},t) = \int_{\rmif}^{\rpif} S_{t-}^{z-1} z e^{\widetilde{\kappa}(z)(T-t)} \,p(z)\,dz,
\end{equation}
which is a \dstr{} if Condition \ref{e:deltaStr2} of Definition \ref{d:deltaStr} is satisfied.
\end{bsp}

\begin{bsp} (\textbf{Mean-variance optimal hedge in the martingale case}) By the integral representation provided in \cite[Theorem~3.1]{hub06}, the mean-variance optimal hedging strategy in an exponential L\'evy model is a \dstr{}, provided that the corresponding asset price is a martingale. In this case, one can therefore use the results of the present paper to quantify the effect of model misspecification arising from using a mean-variance optimal hedge in another model. If the asset price process fails to be a martingale, the corresponding hedge contains a feedback term and therefore is not a \dstr{}. Nevertheless, the results from the martingale case should typically serve as a good proxy, because numerical experiments using \cite[Theorems 3.1 and~3.2]{hub06} supply compelling evidence that the effect of a moderate drift rate is rather small for mean-variance hedging.
\end{bsp}

As stated above, the delta hedge and the mean-variance optimal strategy coincide in the Black-Scholes model. Moreover, the regularity conditions in Definition \ref{d:deltaStr} are automatically satisfied in this case.

\begin{lemma} \label{lem:BS}
Let $\widetilde{S}$ be a geometric Brownian motion without drift, i.e.,
$$\widetilde{S}_t=\widetilde{S}_0e^{-\frac{1}{2}\sigma^2t+\sigma W_t}, \quad \widetilde{S}_0 \in \mathbb{R}_+, \quad t \in [0,T],$$
for a constant $\sigma \in (0,\infty)$ and a standard Brownian motion $W$. Then the delta hedge and the mean-variance optimal hedge in the model~$\widetilde{S}$ coincide and are given by the \dstr{}
$$ \varphi^{BS}_t = \int_{R-i\infty}^{R+i\infty} \widetilde{S}_{t-}^{z-1} ze^{\frac{1}{2}\sigma^2z(z-1)(T-t)}\,p(z)\,dz.$$
\end{lemma}

\begin{prf} We follow the lines of Example~\ref{ex:DeltaHedge} and show that the necessary regularity conditions hold in the Black-Scholes case. First, the cumulant generating function of the driver of~$\widetilde{S}$ is given by
$$\widetilde{\kappa}(z)=\frac{1}{2}\sigma^2z(z-1)$$
for $z\in\C$. Using the arguments of Example~\ref{ex:DeltaHedge}, we obtain the price process of~$H$ in the model~$\widetilde{S}$ as
\begin{equation} \label{eq:BSpriceProc}E\left(f(\widetilde{S}_T)\Big|\F_t\right) = \int_\rmif^\rpif \widetilde{S}^z_t e^{\widetilde{\kappa}(z)(T-t)}\,p(z)\,dz.\end{equation}
Note that the conditional expectation exists and Fubini's Theorem can be applied because all exponential moments of the normal distribution are finite. For the further considerations, note that
$$\Re{\widetilde{\kappa}(z)} = \frac{1}{2}\sigma^2\left(\Re{z}^2-\Re{z}-\Im{z}^2\right),$$
and define $$\widetilde{g}(z,t):=ze^{\widetilde{\kappa}(z)(T-t)}$$
for $z\in\Sf$ and $t\in[0,T]$. Since the mapping $x\mapsto x e^{-ax^2}$, $a>0$, is bounded on $\Rp$,
$$z\mapsto \abs{\widetilde{g}(z,t)} \leq \left(e^{\frac{1}{2}\sigma^2(R^2-R)T}\vee 1\right)\left(\abs{R}+\abs{\Im{z}}e^{-\frac{1}{2}\sigma^2\Im{z}^2(T-t)}\right)$$
is bounded on $R+i\R$ for fixed $t\in[0,T)$ and hence $p$-integrable. Consequently, the integral $\int_\rmif^\rpif s^{z-1}\widetilde{g}(z,t) \,p(z)\,dz$ is well-defined for $s\in (0,\infty)$ and $t\in[0,T)$, and dominated convergence yields
$$\frac{\partial}{\partial s} \int_\rmif^\rpif s^z e^{\widetilde{\kappa}(z)(T-t)}\,p(z)\,dz = \int_\rmif^\rpif s^{z-1} \widetilde{g}(z,t)\,p(z)\,dz.$$
Therefore the delta hedge of the price process in~\eqref{eq:BSpriceProc} is given by
$$\varphi^{\Delta}_t := \int_\rmif^\rpif \widetilde{S}_{t-}^{z-1} \widetilde{g}(z,t)\,p(z)\,dz = \varphi^{BS}_t.$$
Note that whereas this integral is not well-defined for $t=T$, the cumulated gains process $\varphi^{\Delta}\sint \widetilde{S}_T$ does not depend on the value of $\varphi^{\Delta}_T$. Since delta hedging leads to perfect replication in the Black-Scholes model, $\varphi^{\Delta}$ is clearly mean-variance optimal. Let us now verify that $\varphi^{\Delta}$ is indeed a \dstr. Obviously, $\widetilde{g}(z,\cdot)$ is continuous for fixed $z\in\Sf$. Moreover, we have
$$\int_0^T \abs{\widetilde{g}(z,s)}^2 ds \leq \left(e^{\sigma^2(R^2-R)T}\vee 1\right)\left(TR^2 +  \int_0^T \Im{z}^2 e^{-\sigma^2 \Im{z}^2(T-s)}\,ds\right).$$
Elementary integration yields that the right-hand side is uniformly bounded for $z\in\Sf$, which implies that the left-hand side is $p$-integrable. Thus Conditions~\ref{e:deltaStr0} and~\ref{e:deltaStr2} of Definition~\ref{d:deltaStr} are satisfied and we are done.
\end{prf}

\begin{bem} Note that ~$\varphi^{BS}$ is also a replicating strategy and in particular mean-variance optimal for geometric Brownian motion \emph{with} drift, i.e, $$\widetilde{S}_t=\widetilde{S}_0e^{\left(\mu-\frac{1}{2}\sigma^2\right)t+\sigma W_t} \quad \mbox{for } \mu\in (0,\infty).$$
\end{bem}

\begin{bem} \label{rem:BrownComp}
	It can be shown by basically the same arguments as in the proof of Lemma~\ref{lem:BS} that the delta hedge in the sense of Example~\ref{ex:DeltaHedge} exists, admits a representation as in~\eqref{eq:FormulaDeltaHedge} and is a \dstr{} if the risk-neutral driver~$\widetilde{X}$ has a Brownian component.
\end{bem}

% If, in contrast, the driver is a pure-jump process, a sufficiently fast decay of the weight function~$p$ of the payoff also ensures that a delta hedge exists and is a \dstr.
% 
% \begin{lemma} \label{lem:payoffRegularity}
% 	
% \end{lemma}

\section{\texorpdfstring{Performance of a \dstr}{Performance of a Delta-strategy}} \label{sec:PerfDeltaStr}
We measure the performance of a strategy in terms of the resulting mean squared hedging error, i.e., the objective function used in mean-variance hedging.

\begin{defi} \label{def:mshe}
For a given initial endowment $c\in\R$ and a strategy $\varphi\in L(S)$ such that $\varphi \sint S_T\in L^2(P)$, we define the \emph{mean squared hedging error} of the endowment/strategy pair $(c,\varphi)$ as
$$E\left(\left(H - c - \varphi \sint S_T\right)^2\right).$$
\end{defi}

\subsection{Main result}
In this section, we state the main result of this paper. For better readability, the proof is deferred to Appendix~\ref{a:Proof}.

\begin{satz}\label{thm:MainResult}
 Consider an initial endowment~$c\in\R$ and a \dstr~$\varphi$ of the form
 \begin{equation*}%\label{eq:PhiIntegral}
     \varphi_t = \int_{\rmif}^{\rpif} \varphi(z)_t \, p(z)\,dz
 \end{equation*}
 with
 \begin{equation*}%\label{eq:PhiG}
     \varphi(z)_t = S_{t-}^{z-1} g(z,t).
 \end{equation*}
 Then the mean squared hedging error of the endowment/strategy pair $(c,\varphi)$ is given by
 \begin{equation}\label{e:hedgingError}
     E\left(\left(H - c - \varphi \sint S_T\right)^2\right) = (w-c)^2 + \int_{R-i\infty}^{R+i\infty} \int_{R-i\infty}^{R+i\infty} J(y,z) \, p(y)p(z)\, dydz,
 \end{equation}
 where
%     \begin{align}
%           \alpha(z,t) & := \left( 1 - \kappa(1) \int_t^T e^{\kappa(z)(s-T)} g(z,s)\,ds \right) e^{\kappa(z)(T-t)}, \label{e:funcAlpha}  \\[2ex]
%           w & := \int_{\rmif}^{\rpif} S_0^z \alpha(z,0)\,p(z)\,dz,\\[2ex]
%           J(y,z) & :=  \begin{cases}\bigl(\kappa(y+z)-\kappa(y)-\kappa(z)\bigr)\phantom{\,y} \ds\int_0^T S_0^{y+z} e^{\kappa(y+z)s}\alpha(y,s)\alpha(z,s)\,ds  &\\
%             - \,\bigl(\kappa(y+1)-\kappa(y)-\kappa(1)\bigr)\,\ds\int_0^T S_0^{y+z} e^{\kappa(y+z)s} \alpha(y,s) g(z,s)\,ds &\\
%               - \,\bigl(\kappa(z+1)-\kappa(z)-\kappa(1)\bigr)\,\ds\int_0^T S_0^{y+z} e^{\kappa(y+z)s} \alpha(z,s) g(y,s)\,ds & \\
%              +  \,\bigl(\kappa(2)-2\kappa(1)\bigr) \,\ds\int_0^T S_0^{y+z} e^{\kappa(y+z)s} g(y,s) g(z,s)\, ds. \end{cases} \label{e:J}
%     \end{align}
	\begin{align}
       \alpha(z,t) & := \left( 1 - \kappa(1) \int_t^T e^{\kappa(z)(s-T)} g(z,s)\,ds \right) e^{\kappa(z)(T-t)}, \label{e:funcAlpha}  \\[2ex]
       w & := \int_{\rmif}^{\rpif} S_0^z \alpha(z,0)\,p(z)\,dz,
       \end{align}
       and
       \begin{align}
       J(y,z) & :=  \bigl(\kappa(y+z)-\kappa(y)-\kappa(z)\bigr)\phantom{\,y} \ds\int_0^T S_0^{y+z} e^{\kappa(y+z)s}\alpha(y,s)\alpha(z,s)\,ds  \nonumber\\
         &\phantom{:=}\;-\,\bigl(\kappa(y+1)-\kappa(y)-\kappa(1)\bigr)\,\ds\int_0^T S_0^{y+z} e^{\kappa(y+z)s} \alpha(y,s) g(z,s)\,ds \nonumber\\
           &\phantom{:=}\;-\,\bigl(\kappa(z+1)-\kappa(z)-\kappa(1)\bigr)\,\ds\int_0^T S_0^{y+z} e^{\kappa(y+z)s} \alpha(z,s) g(y,s)\,ds  \nonumber\\
          &\phantom{:=}\;+\,\bigl(\kappa(2)-2\kappa(1)\bigr) \,\ds\int_0^T S_0^{y+z} e^{\kappa(y+z)s} g(y,s) g(z,s)\, ds. \label{e:J}
 \end{align}

\end{satz}

\begin{bem}
 The cumulant generating function~$\kappa$ is often known explicitly, e.g., for normal inverse Gaussian \cite{barndorff.98}, variance gamma \cite{madan.senata.90} and CGMY \cite{carr.al.2002} processes or for the  models introduced by Merton \cite{merton.76} and Kou \cite{kou.02}. Moreover, the time integrals in~(\ref{e:funcAlpha}) and in~(\ref{e:J}) can typically be calculated in closed form, which means that the evaluation of the hedging error \eqref{e:hedgingError} usually amounts to numerical integration of a double integral with known integrand as in \cite[Theorem~3.2]{hub06} for the mean-variance optimal hedge.
\end{bem}

\subsection{Sketch of the proof}
Besides giving a rigorous proof of Theorem~\ref{thm:MainResult} in Appendix~\ref{a:Proof}, we present our approach on an intuitive level. The notation used in Appendix~\ref{a:Proof} is anticipated here, but it will be defined precisely in the corresponding places.

To calculate
\begin{equation}\label{e:squHedgingError}
 \ew{\left(H-c-\varphi\sint S_T\right)^2},
\end{equation}
we look for a martingale~$L$ with $L_T = H - c - \varphi\sint S_T$. Then we can rewrite~(\ref{e:squHedgingError}) as
\begin{equation} \label{eq:AnsatzBigL}
 E\left(\left(H-c-\varphi\sint S_T\right)^2\right) = E(L_0^2) + E(\covar{L}{L}_T)
\end{equation}
by means of the predictable quadratic variation of the process~$L$, cf.\ \cite[I.4.2 and~I.4.50(b)]{js03}. Using the integral structure of payoff and strategy, we obtain by a stochastic Fubini argument that
\begin{equation}\label{eq:IdeaFubini}
 H - c - \varphi\sint S_T = \int_{\rmif}^{\rpif} S_T^z \, p(z)\, dz - c - \int_{\rmif}^{\rpif} \left(\varphi(z)\sint S_T\right)\, p(z) \, dz.
\end{equation}
The key idea is to identify a family of martingales~$l(z)$, $z\in\Sf$, such that
$$
 l(z)_T = S_T^z - \varphi(z)\sint S_T.
$$
Then the process
\begin{equation}\label{eq:CandL}
 \int_{\rmif}^{\rpif} l(z)\, p(z) \, dz - c
\end{equation}
is the canonical candidate for the martingale~$L$, and the bilinearity of the predictable covariation~$\covar{\cdot}{\cdot}$ suggests that
\begin{equation}\label{eq:CandCovarL}
\begin{aligned}
 \covar{L}{L} & = \covar{\int_{\rmif}^{\rpif} l(y) \, p(y)\, dy}{\int_{\rmif}^{\rpif} l(z) \, p(z)\, dz} \\
              & = \int_{\rmif}^{\rpif} \int_{\rmif}^{\rpif} \covar{l(y)}{l(z)} \, p(y)p(z) \, dy dz.
\end{aligned}
\end{equation}
In this case, Fubini's Theorem implies
$$
 E\left(\covar{L}{L}_T\right) = \int_{\rmif}^{\rpif} \int_{\rmif}^{\rpif} E\left(\covar{l(y)}{l(z)}_T\right) p(y)p(z) \, dy dz.
$$
We now consider how to determine~$l(z)$. If the stock price $S$ is a martingale, the definition of the cumulant generating function and the martingale property of $\varphi(z) \sint S$ yield that
$$E\left(S_T^z-\varphi(z)\sint S_T|\F_t\right) = S_t^z e^{\kappa(z)(T-t)}-\varphi(z)\sint S_t$$
is the appropriate martingale. Motivated by this fact, we make the ansatz
\begin{equation}\label{e:ansatzL}
 l(z)_t = S^z_t\alpha(z,t) - \varphi(z)\sint S_t
\end{equation}
for deterministic functions $\alpha(z):[0,T]\rightarrow\C$ with $\alpha(z,T)=1$ in the general case.  The drift rate of~(\ref{e:ansatzL}) can be calculated using integration by parts, and setting it to zero yields a linear ODE for the mapping~$t\mapsto\alpha(z,t)$. The solution then leads to the desired candidates for $l(z)$ and $L$ via \eqref{e:ansatzL} resp.\ \eqref{eq:CandL}.

\section{Numerical illustration}\label{s:numillustr}
\begin{figure}[btp]
	\centering
	\subfloat[Mean squared hedging error]{
	\includegraphics[width=0.8\textwidth,keepaspectratio]{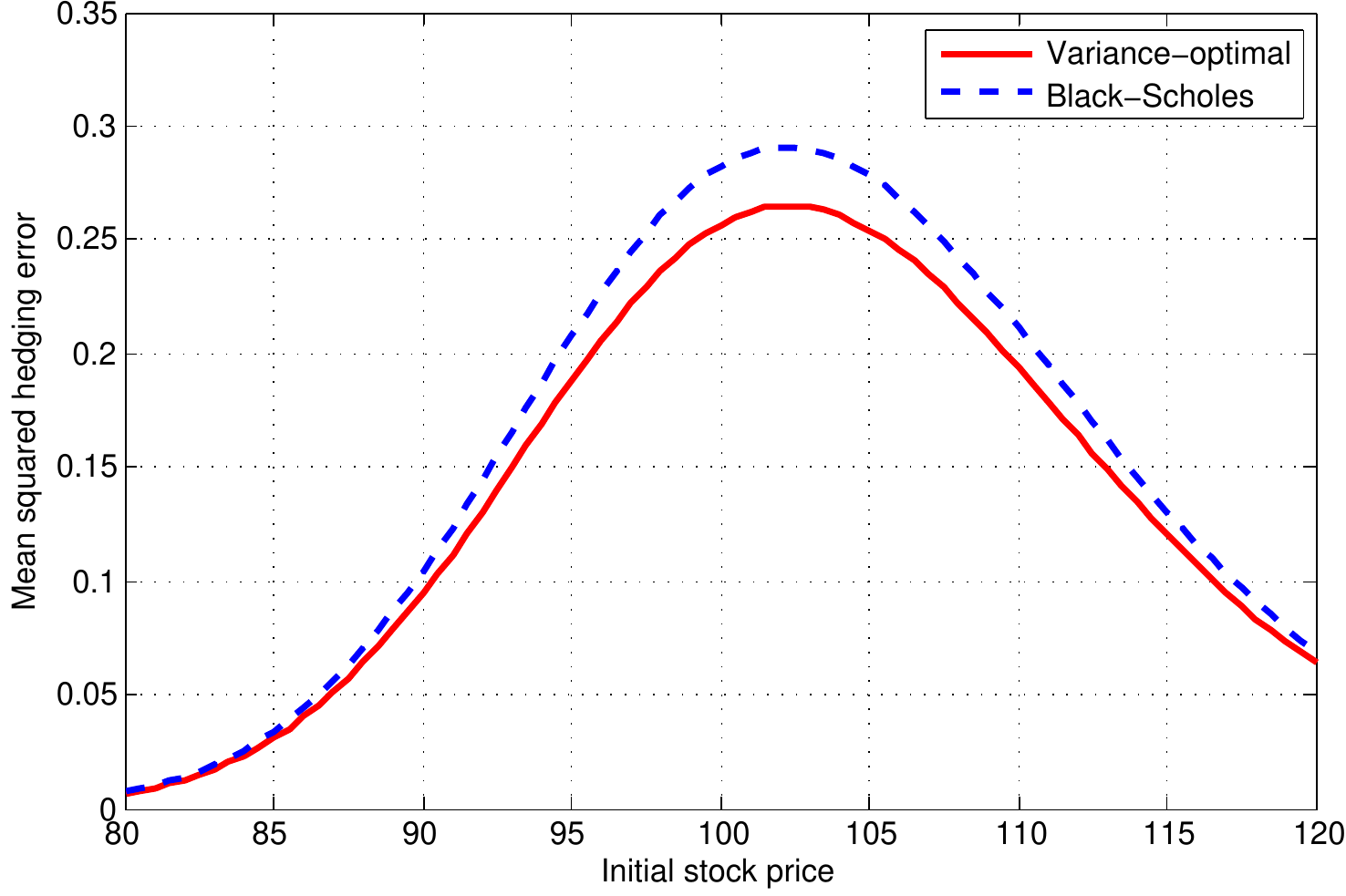}
	%\label{fig:AbsErrVarInitial_NIG}
	}

	\vspace{2cm}

	\subfloat[Relative hedging error]{
	\includegraphics[width=0.8\textwidth,keepaspectratio]{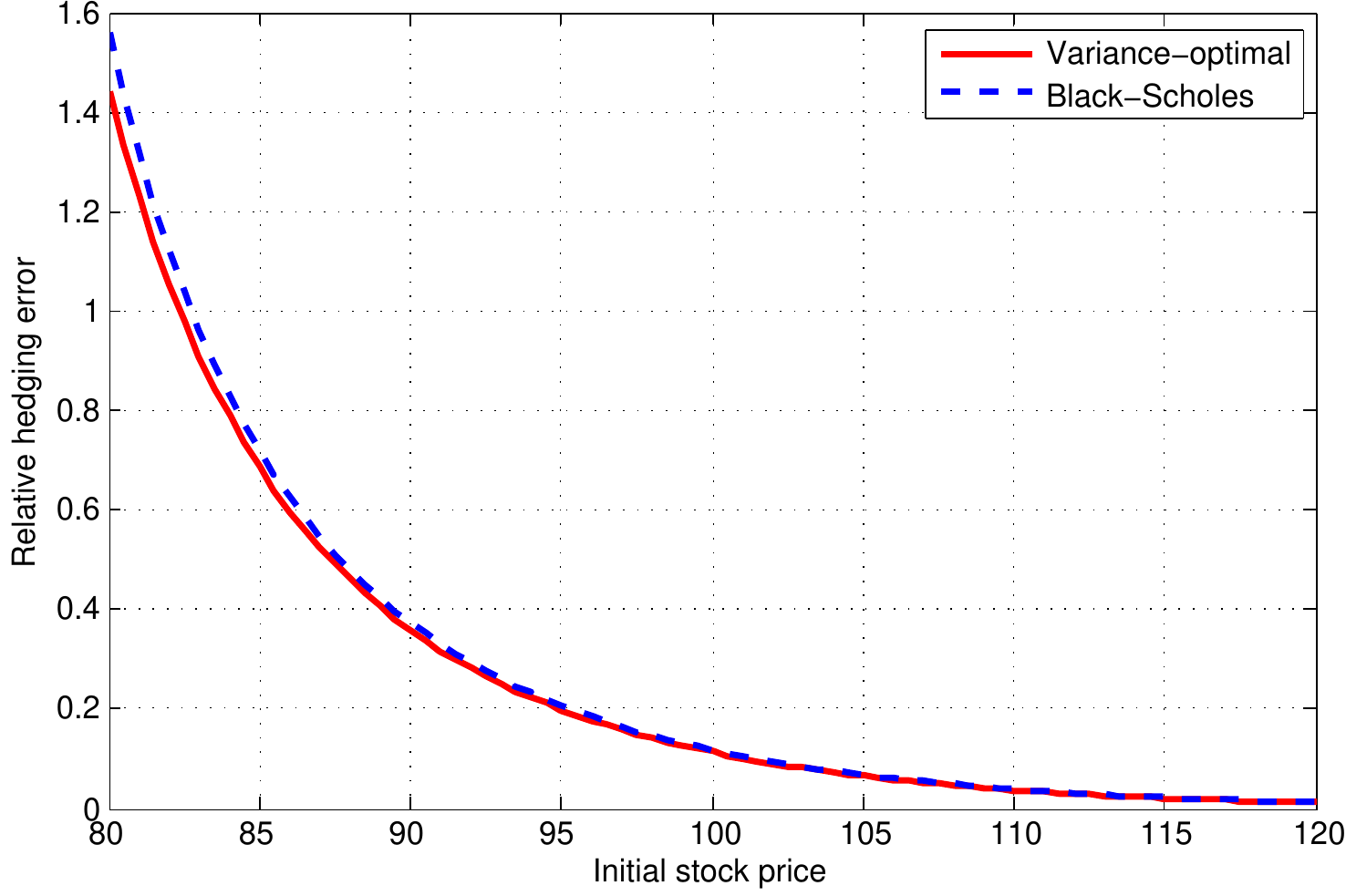}
	%\label{fig:RelErrVarInitial_NIG}
	}

	\vspace{0.5cm}
	\caption{Hedging errorr for varying initial stock price (NIG model under physical measure)}
	\label{fig:ErrVarInitial_NIG}
\end{figure}

\begin{figure}[ttp]
	\begin{center}
		\includegraphics[width=0.8\textwidth,keepaspectratio]{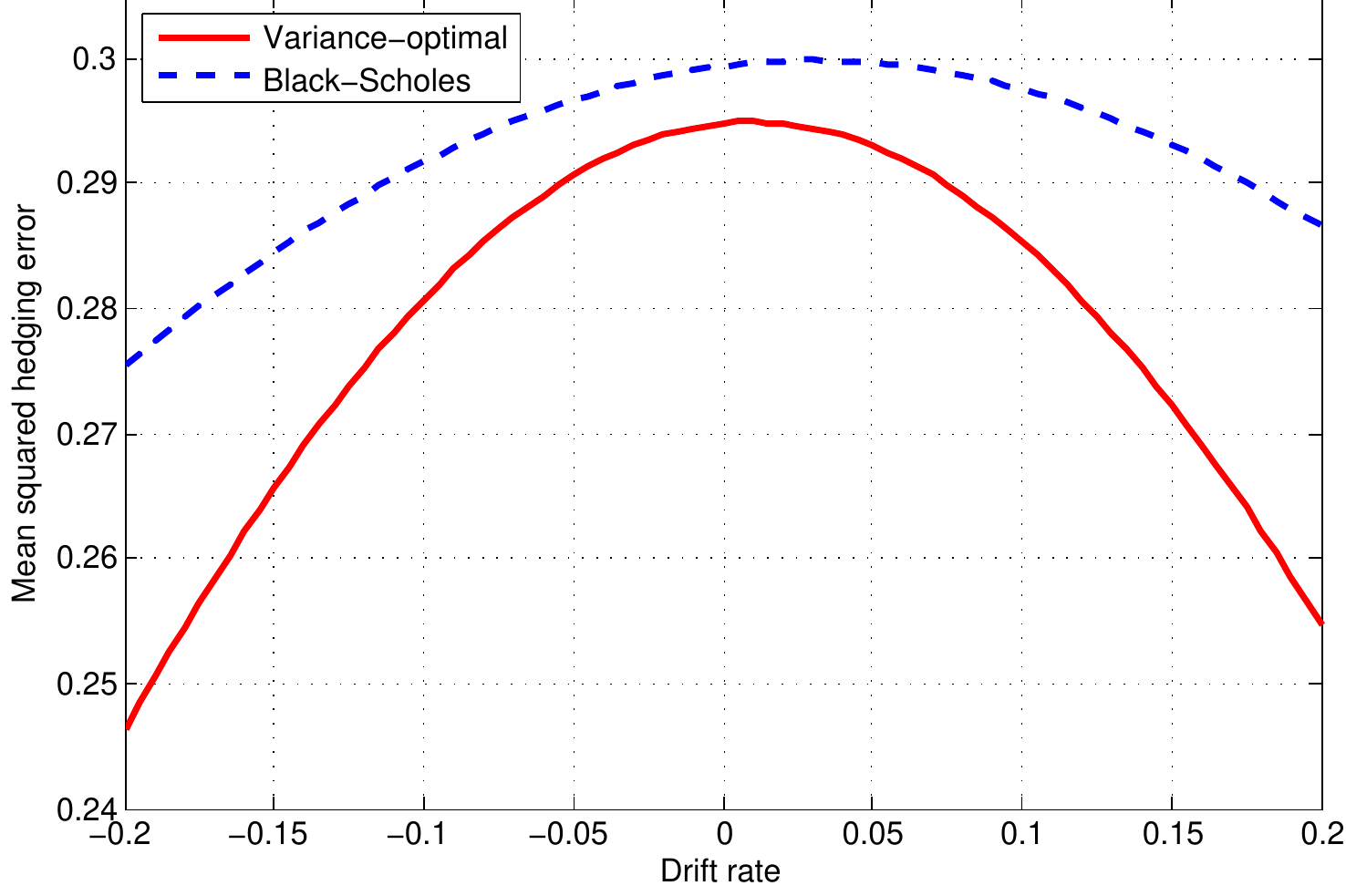}
		\caption{Mean squared hedging error for varying drift rate $\kappa^{\text{NIG}}(1)$ (NIG model under physical measure)} \label{fig:ErrVarDrift}
	\end{center}
\end{figure}

In this section, we illustrate our formulas by two examples. First, we compare the performance of the Black-Scholes strategy and the variance-optimal strategy in the \emph{normal inverse Gaussian} (henceforth NIG)  \levy{} model (cf.\ \cite{barndorff.98}) with parameters inferred from a historical time series, i.e., we use the physical probability measure. Afterwards, we assess the hedging errors of the Black-Scholes strategy, the delta hedge and the variance-optimal strategy in the \emph{diffusion-extended CGMY} (CGMYe) model (cf.\ \cite{carr.al.2002}) for parameters obtained by a calibration to market prices of options, i.e., in this case we use a risk-neutral probability measure. To evaluate the mean-variance optimal hedging error and the corresponding optimal initial capital, we use the formulas of \cite{hub06}.

We assume a riskless interest rate of~4\% and consider a European call option with maturity $T=0.25$ years and discounted strike $K=99$. The integral representation of the corresponding payoff function is given in Example~\ref{ex:Call}.

For the numerical evaluation of the integrals we use the routine \emph{{gsl\_integration\_qagi}} from the quadrature framework of the \emph{GNU Scientific Library} (GSL, cf.\ \cite{gslref08}), which computes integrals on the whole real line by a transformation on $[-1,0) \cup (0, 1]$ and then applies an adaptive algorithm using a 15-point Gau\ss-Kronrod scheme. For the parameter $R$ we choose $R=1.1$.  The outcome of the numerical computations is quite robust with respect to the choice of $R$; instabilities occur only if $R$ is chosen very close to or very far away from~$1$.

\subsection{Physical measure}\label{ss:NIG}

In the following example, we examine the quality of the Black-Scholes strategy as a proxy for the variance-optimal hedge in the NIG model. Recall that by Lemma~\ref{lem:BS}, the Black-Scholes strategy is given by the replicating Black-Scholes delta using the volatility parameter~$\sigma$ of the physical price process dynamics, i.e., no risk-neutral (pricing) measure is necessary to determine the hedging strategy in this case.

The cumulant generating function of the NIG \levy{} process is given by \[ \kappa^{\text{NIG}}(z) = \mu z + \delta \left( \sqrt{\alpha^2-\beta^2} - \sqrt{\alpha^2-(\beta+z)^2} \right) \] for $\mu\in\R$, $\delta>0$, $0\leq \abs{\beta}\leq \alpha$ and $z\in  \left\{ y\in\C : |\beta+\Re{y}| \leq \alpha \right\}$ (cf., e.g., \cite[Section~5.3.2]{hub06}). As for parameters, we use \[\alpha=75.49,\quad \beta=-4.089,\quad \delta=3.024,\quad \mu=-0.04,\] which corresponds to the annualized daily estimates from \cite{rydberg97} for a historical time series of Deutsche Bank, assuming 252 trading days per year. It is easily verified that this market satisfies the prerequisites of Section~\ref{s:Model}. The volatility parameter $\sigma$ for the Black-Scholes strategy (cf.\ Lemma~\ref{lem:BS}) is set to $0.2$ so that the log-returns in the corresponding Black-Scholes market and in the NIG \levy{} market exhibit the same variance.

Figure~\ref{fig:ErrVarInitial_NIG} shows the mean squared hedging error and the relative hedging error (i.e., the root of the mean squared hedging error divided by the initial capital) of the Black-Scholes strategy and the mean-variance optimal strategy for varying initial stock price. As initial capital for the Black-Scholes hedge we use the Black-Scholes price of the option, which is virtually indistinguishable from the variance-optimal initial capital, though (compare \cite{hub06}). For an at-the-money call the relative hedging errors differ by 4.73\% and amount to 0.113 (mean-variance optimal) and 0.118 (Black-Scholes).

Figure~\ref{fig:ErrVarDrift} illustrates how the two strategies react to different drift rates of the underlying. More specifically, the figure shows the mean squared hedging error for an at-the-money call option with strike $K=100$ and maturity $T=0.25$ for varying drift rate $\kappa(1)$, controlled by varying the location parameter $\mu$ of the NIG process. Since the Black-Scholes strategy does not incorporate such systematic drifts directly, the two errors differ least in the martingale case $\kappa(1)=0$. Altogether, the Black-Scholes strategy seems to be a surprisingly good proxy for the mean-variance optimal hedge, particularly for moderate drift rates.

\subsection{Risk-neutral measure} \label{ss:CGMYE}
In this second example, we compare the performance of the Black-Scholes hedge, the delta hedge and the variance-optimal hedge in the CGMYe model. The determination of the delta hedge and the variance-optimal hedge as well as the computation of the hedging error all take place with respect to a risk-neutral measure inferred from a calibration of the model to option prices. The volatility parameter~$\sigma$ for the Black-Scholes hedge is chosen for each initial stock price such that the resulting Black-Scholes price of the option matches the one implied by the CGMYe model.

Let us emphasize that one has to be careful with the interpretation of hedging errors computed under a risk-neutral probability measure. This is because expected values under this measure have no direct statistical meaning. Nevertheless, this approach is quite common in the literature, cf., e.g., \cite{cont.al.05} and the references therein. For a more detailed empirical investigation taking care of this issue, one would first determine the delta hedge under a risk-neutral measure obtained by calibration. In a second step, one would then estimate a parametric ansatz for the market price of risk in order to switch to an appropriate physical measure. However, this is beyond our scope here and is therefore left to future research.

The cumulant generating function of the risk-neutral CGMYe \levy{} process is given by $$\kappa^{\text{CGMYe}}(z) = \omega-\frac{1}{2}\eta^2 z + \frac{1}{2}\eta^2z^2 + C\Gamma(-Y)\left( (M-z)^Y - M^Y + (G+z)^Y - G^Y\right),$$
where $\Gamma$ denotes the Gamma function and $C>0$, $G>0$, $M\geq 0$, $Y<2$ and $\eta>0$ are the parameters of the model. Given these quantities, the value of $\omega$ is chosen such that $\kappa^{\text{CGMYe}}(1)=0$, which implies that the stock price process is a martingale. For a more detailed discussion of the meaning of the different parameters we refer to \cite{carr.al.2002}. Here, we use the values from the same paper given by $$C = 9.61,\quad G = 9.97,\quad M = 16.51,\quad Y = 0.1430,\quad \eta = 0.0458.$$

Note that since $\eta > 0$, the driving \levy{} process has a Brownian component and hence the delta hedge exists and is a \dstr{} by Remark~\ref{rem:BrownComp}. Moreover, the model satisfies the prerequisites of Section~\ref{s:Model}.

\begin{figure}[tbp]
	\centering
	\subfloat[Mean squared hedging error]{
	\includegraphics[width=0.8\textwidth,keepaspectratio]{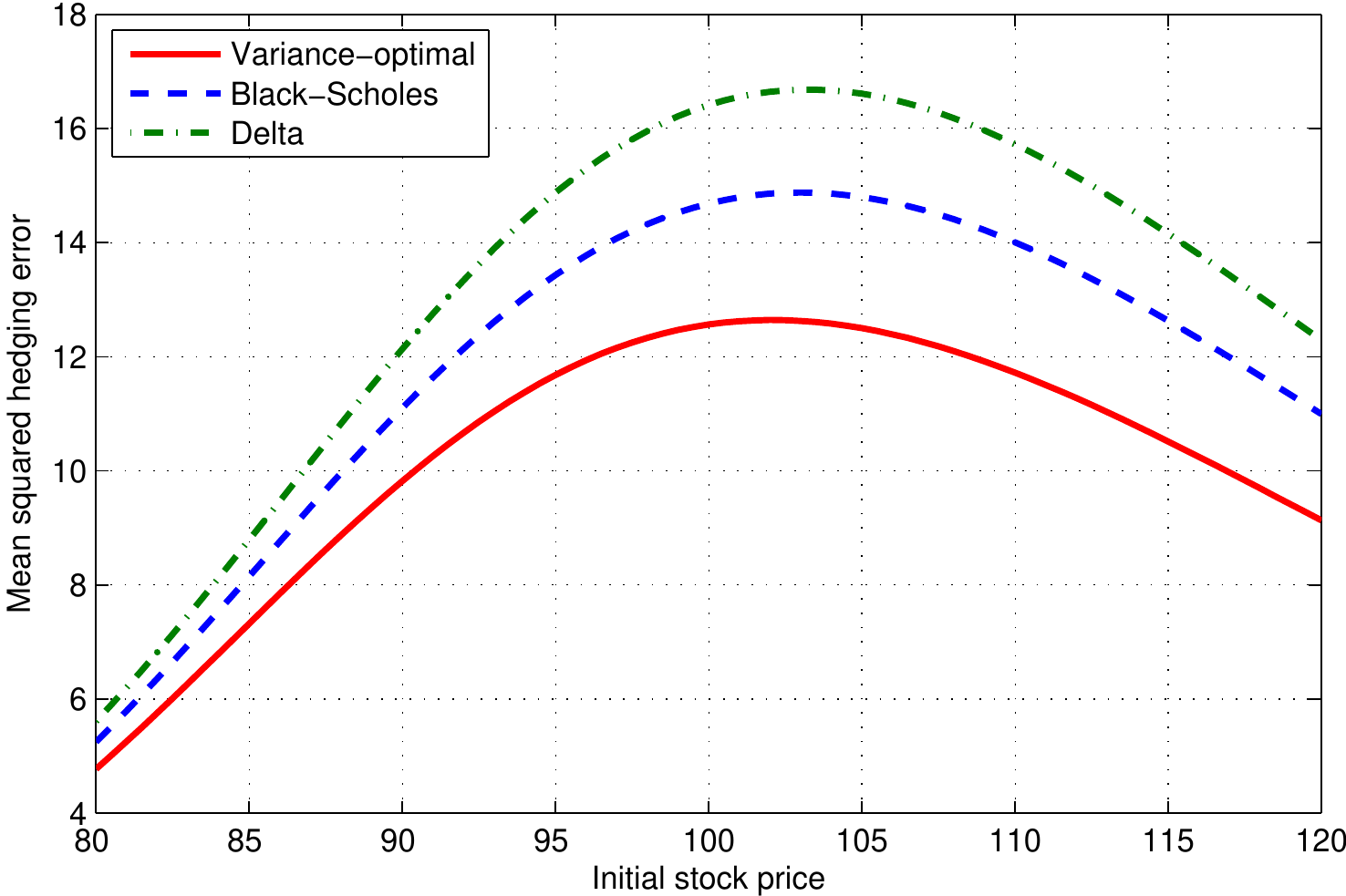}
	%\label{fig:AbsErrVarInitial_CGMY}
	}
	
	\vspace{2cm}
	
	\subfloat[Relative hedging error]{
	\includegraphics[width=0.8\textwidth,keepaspectratio]{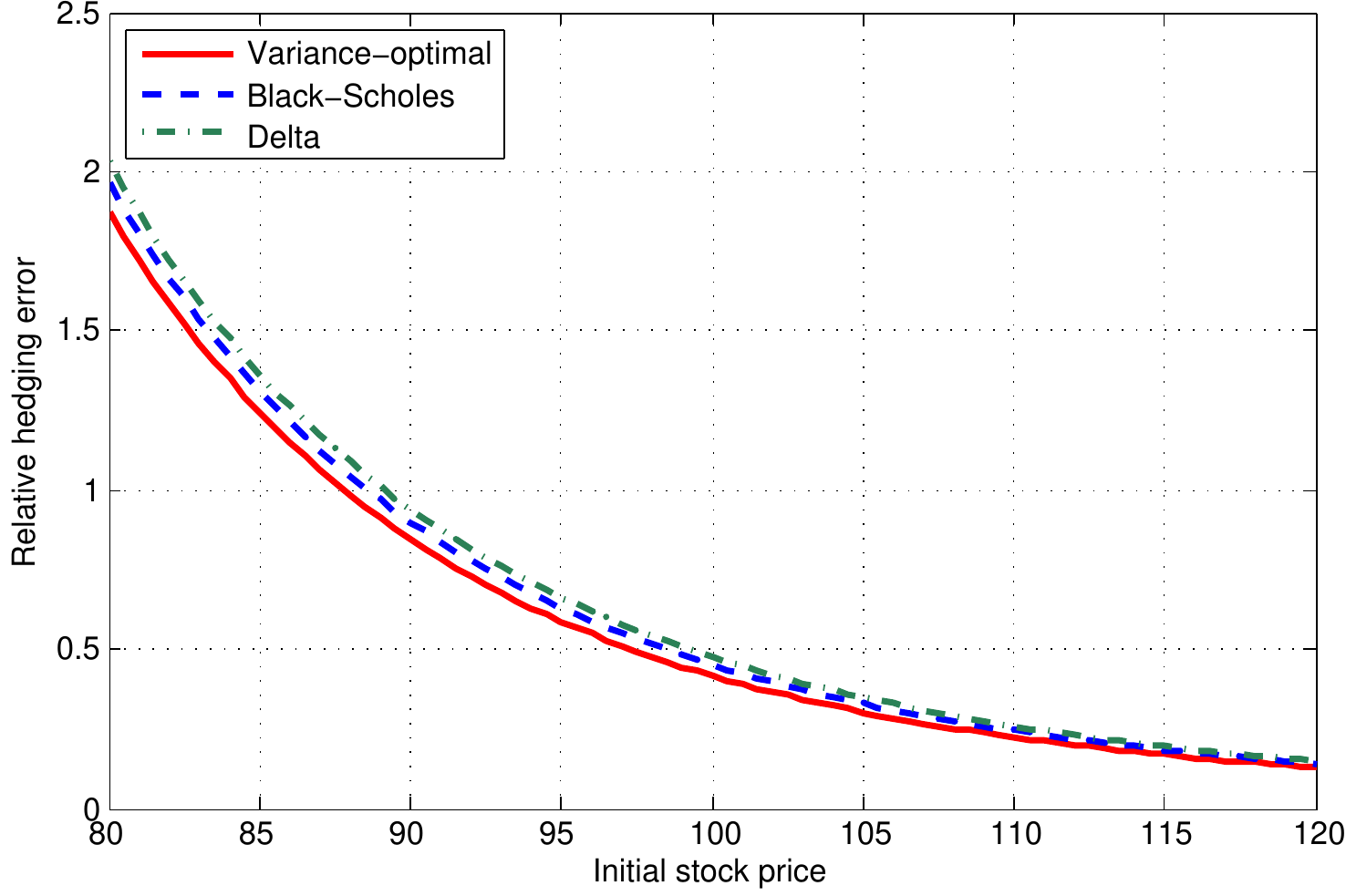}
	%\label{fig:RelErrVarInitial_CGMY}
	}

	\vspace{0.5cm}
	\caption{Hedging errors for varying initial stock price (CGMYe model under risk-neutral measure)}
	\label{fig:ErrVarInitial_CGMY}
\end{figure}

Figure~\ref{fig:ErrVarInitial_CGMY} shows the mean squared hedging error and the corresponding relative hedging error (cf.\ Section~\ref{ss:NIG}) of the variance-optimal strategy, the Black-Scholes and the delta hedge for different initial stock prices in the CGMYe model with parameters as explained above. The mean squared errors for an at-the-money option amount to 12.57 for the variance-optimal hedge, to 14.68 for the Black-Scholes hedge and to 16.41 for the delta hedge. This corresponds to a deviation of the relative errors of 8.10\% (Black-Scholes) and 14.3\% (delta) from the variance-optimal value.

In contrast to the example in Section~\ref{ss:NIG}, mean squared and relative hedging errors in general are much higher and the Black-Scholes strategy performs, compared to the variance-optimal one, worse than in the previous example. This can be explained by the fact that skewness and especially excess kurtosis of the driving L\'evy process are much more pronounced in this case. Indeed, in the CGMYe model skewness and excess kurtosis of the daily logarithmic returns amount to -3.852 and 62.32 resp.\ to -0.2384 and 0.2416 for the yearly logarithmic returns. In contrast, skewness and excess kurtosis in the NIG model are given by -0.1709 and 3.356 for the daily logarithmic returns and by -0.0108 and 0.0133 for the yearly returns. Moreover, the mean squared and relative hedging error of the Black-Scholes strategy is considerably lower than that of the delta hedge. This effect has also been observed by \cite{angelini.herzel.10} for stochastic volatility models in a discrete time setup.

Altogether, despite the much higher skewness and kurtosis, the Black-Scholes hedge still seems to be a quite stable proxy for variance-optimal one and, in particular, performs noticeably better than the model delta.

\section{Conclusion}
In general exponential L\'evy models, we have derived semi-explicit formulas for the mean-squared hedging error of a European-style contingent claim. This has been done for so-called $\Delta$-strategies, which include the Black-Scholes hedging strategy and more general delta hedges. Numerical examples obtained by implementing these results show that -- both under the physical and subject to a risk-neutral probability -- the Black-Scholes hedge seems to perform surprisingly well also in L\'evy models with jumps. Nevertheless, it does lead to a non-trivial hedging error in this case, which can be quantified using our approach.

% \begin{figure}[htbp]
%   \centering
%   \subfigure[Mean squared hedging error for varying initial stock price]{
%     \label{fig:ErrVarInitial}
%     \includegraphics[width=0.9\textwidth]{hedging_error}
%   }
%   \subfigure[Mean squared hedging error for varying drift rate $\kappa(1)$]{
%     \label{fig:ErrVarDrift}
%     \includegraphics[width=0.9\textwidth]{hedging_error_drift}
%   }
% \end{figure}

\begin{appendix}
\section{Proof of the main result}{\label{a:Proof}}
In this appendix, we present the proof of Theorem~\ref{thm:MainResult}, which is split up into several intermediate statements. An essential tool for the forthcoming considerations are the special semimartingale decomposition and the predictable covariation of complex powers of~$S$, provided by the following
\begin{lemma}
 \label{l:DecompSpowY}
  For $z\in\Sf$, the process $S^z$ is a special semimartingale whose canonical decomposition $S^z = S_0^z + M(z)+A(z)$ is given by
 \begin{equation*}
     M(z)_t  = \ds\int_0^t e^{\kappa(z)s}\,dN(z)_s, \quad A(z)_t = \kappa(z) \ds\int_0^t S_{s-}^z\,ds,
 \end{equation*}
 where
 \begin{equation*}
     N(z)_t := e^{-\kappa(z)t}S_t^z.
 \end{equation*}
 Moreover, for $y,z\in\Sf$ and continuously differentiable functions $\beta, \gamma:[0, T]\rightarrow \C$, the process $[S^y\beta,S^z\gamma]$ is a special semimartingale with compensator $\covar{S^y\beta}{S^z\gamma}$ given by
 \begin{equation} \label{eq:CovarSySz}
     \covar{S^y\beta}{S^z\gamma}_t = \left(\kappa(y+z)-\kappa(y)-\kappa(z)\right) \ds\int_0^t S_{s-}^{y+z}\beta_s\gamma_s\,ds.
 \end{equation}
\end{lemma}
\begin{prf}
 This follows along the lines of the proof of \cite[Lemma~3.2]{hub06}.
\end{prf}

With the special semimartingale composition of $S$ at hand, we can now establish that the mean squared hedging error of a \dstr{} is well-defined.

\begin{lemma}\label{l:L(S)}
The \dstr{} $\varphi$ satisfies $\varphi \in L(S)$ and $\varphi \sint S_T \in L^2(P)$.
\end{lemma}

\begin{prf}
Fubini's Theorem yields the predictability of $\varphi$. The assertion then follows from \cite[Lemma~3.1]{hub06}, Fubini's Theorem, H\"older's inequality and Condition 2 of Definition \ref{d:deltaStr} since $t \mapsto E(|S^{z}_t|^2)=S_0^{2R}e^{t\kappa(2R)}$ is bounded on [0,T] for $z \in\Sf$  by \eqref{e:secMomRX}.
\end{prf}

The following proposition ascertains that deterministic integration in the representation of $\varphi$ and stochastic integration with respect to~$S$ can be interchanged, compare \eqref{eq:IdeaFubini}.
\begin{prop} \label{p:fubini}
 We have
 \begin{equation*}
     \varphi\sint S = \int_{\rmif}^{\rpif} \left( \varphi(z)\sint S \right) p(z)\,dz.
 \end{equation*}
\end{prop}

\begin{prf}
By the definition of $L(S)$ in \cite[III.6.6(c)]{js03}, Lemma \ref{l:DecompSpowY} and since $E(S^2_{t-})=S_0^2e^{t\kappa(2)}$ and the locally bounded process $S_{-}$ are bounded resp.\ pathwise bounded on $[0,T]$, a deterministic process $(H_t)_{t \in [0,T]}$ belongs to $L(S)$ if $\int_0^T H^2_t dt <\infty$.  Hence Condition 2 of Definition \ref{d:deltaStr} implies in combination with Fubini's Theorem that
$$\left(\int_{\rmif}^{\rpif} |g(z,\cdot)|^2 |p(z)|\,|dz| \right)^{1/2} \in L(S).$$
Since $S_{-}^{R-1}$ is locally bounded, \cite[III.6.6.19(e)]{js03} yields
 \begin{equation*}
      \left(\int_{\rmif}^{\rpif} |\varphi(z)|^2 |p(z)|\,|dz|\right)^{1/2} = S_{-}^{R-1} \left(\int_{\rmif}^{\rpif} |g(z,\cdot)|^2 |p(z)|\,|dz| \right)^{1/2} \in L(S).
 \end{equation*}
The assertion now follows from Fubini's Theorem for stochastic integrals, cf., e.g., \cite[Theorems~63 and~65]{protter04}.
\end{prf}

The following theorem shows that our Ansatz~\eqref{e:ansatzL} indeed works.

\begin{satz}
 \label{thmCorrectionErrorFixedZ}
 Let $z\in\Sf$ and define $\alpha(z):[0,T]\rightarrow\C$ as
 \begin{equation} \label{sol_Alpha}
      \alpha(z,t) := \left( 1 - \kappa(1) \int_t^T e^{\kappa(z)(s-T)} g(z,s)\,ds \right) e^{\kappa(z)(T-t)}.
 \end{equation}
 Then $\alpha$ solves the terminal value problem
 \begin{equation}
     \label{thmCorrectionErrorFixedZ_ODE}
     \left\{
     \begin{array}{lcl}
          \alpha'(z,t) + \kappa(z)\alpha(z,t) - \kappa(1)g(z,t) & = & 0, \quad  t \in [0,T], \\[1ex]
         \alpha(z,T) & = & 1,
     \end{array}
     \right.
 \end{equation}
 and the process $l(z)$ defined by
 \begin{equation} \label{defn_zErrorProc}
     l(z)_t := S_t^z\alpha(z,t) - \varphi(z)\sint S_t
 \end{equation}
 is a local martingale with $l(z)_T = S_T^z - \varphi(z)\sint S_T$.
\end{satz}

\begin{prf}
 Let $z\in\Sf$. Differentiation shows that $\alpha$ solves~(\ref{thmCorrectionErrorFixedZ_ODE}). To prove the second part of the assertion, we decompose $l(z)$ into a local martingale and a drift, and then conclude that the latter vanishes due to the choice of~$\alpha$. Integration by parts and \cite[I.4.49(d)]{js03} yield
 $$
     l(z) = S_0^z\alpha(z,0) + \int_0^\cdot S_{s-}^z\alpha'(z,s)\,ds + \alpha(z,\cdot) \sint S^z - \varphi(z) \sint S.
 $$
 It now follows from Lemma~\ref{l:DecompSpowY} that
 \begin{gather} \label{eq_Decomp_lz}
     \begin{aligned}
         l(z) = & \; S_0^z\alpha(z,0) + \ds\int_0^\cdot S_{s-}^z\alpha'(z,s)\,ds + \alpha(z,\cdot) \sint A(z) + \alpha(z,\cdot) \sint M(z) \\
         & - \varphi(z)\sint A(1) - \varphi(z) \sint M(1) \\[1.5ex]
          = & \; S_0^z\alpha(z,0) + \alpha(z,\cdot)\sint M(z) - \varphi(z) \sint M(1) \\
         & +\ds\int_0^\cdot S_{s-}^z \Big(\alpha'(z,s) + \kappa(z)\alpha(z,s) - \kappa(1)g(z,s)\Big)\,ds\,.
     \end{aligned}
 \end{gather}
 Since $\alpha$ satisfies~(\ref{thmCorrectionErrorFixedZ_ODE}), the last integral on the right-hand side of~(\ref{eq_Decomp_lz}) vanishes. The remaining terms are local martingales by \cite[I.4.34(b)]{js03}, because $\alpha(z,\cdot)$ and $\varphi(z)$ are locally bounded. Since $\alpha(z,T)=1$, this proves the second part of the assertion.
\end{prf}

Lemma \ref{l:DecompSpowY} now allows us to compute the predictable quadratic covariations $\langle l(y),l(z)\rangle$.

\begin{prop}\label{p:predCovarL}
 \label{propPredCovarl}
 For all $y,z\in\Sf$, the process $[l(y),l(z)]$ is a special semimartingale with compensator $\covar{l(y)}{l(z)}$ given by
%     \begin{equation}\label{e:predCovarL}
%          \covar{l(y)}{l(z)}_t =  \begin{cases}\bigl(\kappa(y+z)-\kappa(y)-\kappa(z)\bigr)\phantom{\,y} \ds\int_0^t S_{s-}^{y+z}\alpha(y,s)\alpha(z,s)\,ds  &\\
%             - \,\bigl(\kappa(y+1)-\kappa(y)-\kappa(1)\bigr)\,\ds\int_0^t S_{s-}^{y+z} \alpha(y,s) g(z,s)\,ds &\\
%               - \,\bigl(\kappa(z+1)-\kappa(z)-\kappa(1)\bigr)\,\ds\int_0^t S_{s-}^{y+z} \alpha(z,s) g(y,s)\,ds & \\
%              +  \,\bigl(\kappa(2)-2\kappa(1)\bigr) \,\ds\int_0^t S_{s-}^{y+z} g(y,s) g(z,s)\, ds. \end{cases}
%     \end{equation}
	\begin{align}
      \covar{l(y)}{l(z)}_t & =  \bigl(\kappa(y+z)-\kappa(y)-\kappa(z)\bigr)\phantom{\,y} \ds\int_0^t S_{s-}^{y+z}\alpha(y,s)\alpha(z,s)\,ds  \nonumber\\
         &\phantom{=}\;- \,\bigl(\kappa(y+1)-\kappa(y)-\kappa(1)\bigr)\,\ds\int_0^t S_{s-}^{y+z} \alpha(y,s) g(z,s)\,ds \nonumber\\
           &\phantom{=}\;- \,\bigl(\kappa(z+1)-\kappa(z)-\kappa(1)\bigr)\,\ds\int_0^t S_{s-}^{y+z} \alpha(z,s) g(y,s)\,ds  \nonumber\\
          &\phantom{=}\;+  \,\bigl(\kappa(2)-2\kappa(1)\bigr) \,\ds\int_0^t S_{s-}^{y+z} g(y,s) g(z,s)\, ds. \label{e:predCovarL}
 	\end{align}

\end{prop}

\begin{prf}
 Let $y, z\in\Sf$. By the definition of $l(\cdot)$ in~(\ref{defn_zErrorProc}), the bilinearity of the quadratic covariation~$[\cdot,\cdot]$ and \cite[I.4.54]{js03}, we obtain that
 \begin{align*}
     [l(y), l(z)] = & \;  [S^y\alpha(y),S^z\alpha(z)] -  \varphi(z)\sint[S^y\alpha(y), S] \\
        & \;- \varphi(y)\sint[S, S^z\alpha(z)] + (\varphi(y)\varphi(z))\sint[S, S],
 \end{align*}
 because $\varphi(y)$ and $\varphi(z)$ are locally bounded. Recall that by Lemma~\ref{l:DecompSpowY}, the square bracket processes on the right-hand side are special semimartingales with compensators given by \eqref{eq:CovarSySz}. Again using that $\varphi(z)$ is locally bounded, it then follows from \cite[I.4.34(b)]{js03} that
 \begin{equation*}
     %\label{eq_propPredCovarl_1}
     \begin{aligned}
         {[l(y),l(z)]} &-\covar{S^y\alpha(y)}{S^z\alpha(z)}-\varphi(z)\sint\covar{S^y\alpha(y)}{S} \\
         & \quad -\varphi(y)\sint\covar{S}{S^z\alpha(z)}+(\varphi(y)\varphi(z))\sint \covar{S}{S}
     \end{aligned}
 \end{equation*}
 is a local martingale. By inserting the explicit representations~\eqref{eq:CovarSySz}, we obtain that our candidate for $\covar{l(y)}{l(z)}$ indeed compensates $[l(y),l(z)]$. Since it is also predictable and of finite variation, this completes the proof.
\end{prf}

The following technical lemma provides the $p$-integrability for several expressions, which is necessary to apply Fubini arguments in the proofs of Propositions~\ref{p:L} and \ref{p:CovarL}.

% \begin{lemma} \label{l:bounds}
%     There exist constants $b_1, b_2, b_3\geq 0$ and, for almost all $\omega\in\Omega$, $b_4(\omega) \geq 0$ such that
%     \begin{enumerate}
%       \item $\Re{\kappa(z)} \leq b_1$, \label{e:bndKappa}
%       \item $|\alpha(z,t)| \leq b_2$, \label{e:bndAlpha}
%       \item $\ew{\left|\covar{l(y)}{l(z)}_t\right|} \leq b_3$, \label{l:boundEx}
%       \item $\left|\covar{l(y)}{l(z)}_t(\omega) \right| \leq b_4(\omega)$, \label{e:boundCoverOmega}
%     \end{enumerate}
%     for all $y,z\in\Sf$ and $t\in[0,T]$, where~$\alpha$ and~$l(z)$ are defined as in~(\ref{sol_Alpha}) and~(\ref{defn_zErrorProc}).
% \end{lemma}

\begin{lemma} \label{l:bounds}
 With $\alpha$ and $l(z)$ defined as in~(\ref{sol_Alpha}) and~(\ref{defn_zErrorProc}), we have the following for all $t\in[0,T]$:
 \begin{enumerate}
   \item There exists a constant $b_1$ such that $\Re{\kappa(z)} \leq b_1$ for all $z\in R+i\R$. \label{e:bndKappa}
   \item The mappings $z\mapsto\alpha(z,t)$ and $z\mapsto\alpha(z,t)^2$ are $p$-integrable. \label{e:bndAlpha}
   \item The mapping $(y,z)\mapsto\ew{\left|\covar{l(y)}{l(z)}_t\right|}$ is twice $p$-integrable. \label{l:boundEx}
   \item The mapping $(y,z)\mapsto\covar{l(y)}{l(z)}_t(\omega)$ is twice $p$-integrable for almost all $\omega\in\Omega$. \label{e:boundCoverOmega}
 \end{enumerate}
\end{lemma}

\begin{prf}
 Let $y,z\in\Sf$ and $t\in[0,T]$.
\begin{enumerate}[leftmargin=*]
\item The definition of the cumulant generating function and Jensen's inequality yield
\begin{equation*}
    e^{\Re{\kappa(z)}} = \left|\ew{e^{zX_1}}\right| \leq \ew{\left|e^{zX_1}\right|} = \ew{e^{\Re{z}X_1}} = e^{\kappa(\Re{z})} = e^{\kappa(R)} \leq e^{b_1}
\end{equation*}
for $b_1 := \kappa(R) \vee 0$.
\item By H\"older's inequality, we have
\begin{align*}
    \left| \alpha(z,t) \right| & \leq e^{\Re{\kappa(z)}(T-t)} + \left|\kappa(1)\right| \int_t^T e^{\Re{\kappa(z)}(s-t)}\left|g(z,s)\right|\,ds \\
     & \leq e^{b_1T} + \left|\kappa(1)\right|e^{b_1(T-t)} (T-t)^{\frac{1}{2}} \left( \int_t^T \left|g(z,s)\right|^2\,ds \right)^{1/2}.
\end{align*}
The $p$-integrability of $z\mapsto\alpha(z,t)$ now follows, because $z\mapsto 1$ ist $p$-integrable by Assumption~\ref{a:intReprF} and the last integral on the right-hand side is $p$-integrable by H\"older's inequality and the fact that Condition~\ref{e:deltaStr2} in Definition~\ref{d:deltaStr} holds for the \dstr{} $\varphi$. Considering the square of the right-hand side, it follows directly from Assumption~\ref{a:intReprF} and Condition~\ref{e:deltaStr2} in Definition~\ref{d:deltaStr} that $z\mapsto\alpha(z,t)^2$ is $p$-integrable.
\end{enumerate}
For the proof of Assertions~\ref{l:boundEx} and~\ref{e:boundCoverOmega}, first note that the bilinearity of $\langle \cdot, \cdot \rangle$ yields
\begin{equation}\label{eq:decompdCovarL}
\pm \Re{\covar{l(y)}{l(z)}} = \frac{1}{2} \Big(\covar{l(y) \pm \overline{l(z)}}{\overline{l(y) \pm \overline{l(z)}}} - \covar{l(y)}{\overline{l(y)}} - \covar{l(z)}{\overline{l(z)}}\Big)
\end{equation}
and
\begin{equation*}
\label{eq_propPredCovarlBounded_3}
\begin{aligned}
    \covar{l(y) \pm \overline{l(z)}}{\overline{l(y) \pm \overline{l(z)}}} & \leq \covar{l(y) \pm \overline{l(z)}}{\overline{l(y) \pm \overline{l(z)}}} + \covar{l(y) \mp \overline{l(z)}}{\overline{l(y) \mp \overline{l(z)}}} \\
    & =  2 \Big(\covar{l(y)}{\overline{l(y)}} + \covar{l(z)}{\overline{l(z)}}\Big).
\end{aligned}
\end{equation*}
Applying an analogous polarization argument to $\Im{\covar{l(y)}{l(z)}}$ by replacing $\overline{l(z)}$ with $i\overline{l(z)}$, we see that it is sufficient to consider only the covariation of the form $\covar{l(z)}{\overline{l(z)}}$ in order to show~\ref{l:boundEx} and~\ref{e:boundCoverOmega}. Since this process is real-valued and increasing, we can restrict ourselves to $t=T$.
\begin{enumerate}[leftmargin=*]
\setcounter{enumi}{2}
\item By applying the arguments from the proof of Proposition~\ref{p:predCovarL} to $\left[l(z),\overline{l(z)}\right]$ instead of $\left[l(y),l(z)\right]$ and taking absolute values, we obtain
\begin{equation}\label{e:PboundsE1}
\begin{aligned}
\covar{l(z)}{\overline{l(z)}}_T & \leq \int_0^T S_{s-}^{2\Re{z}} \abs{\alpha(z,s)}^2 \abs{\kappa(2\Re{z})-2\Re{\kappa(z)}}\,ds \\
& \quad + \int_0^T S_{s-}^{2\Re{z}} \abs{g(z,s)}^2 \abs{\kappa(2)-2\kappa(1)} \,ds \\
& \quad + \int_0^T 2 S_{s-}^{2\Re{z}} \abs{\alpha(z,s)} \abs{g(z,s)} \abs{\kappa(z+1)-\kappa(z)-\kappa(1)} \,ds.
\end{aligned}
\end{equation}
Using
\[ E\left(S_{s-}^{2\Re{z}}\right) = E\left(S_s^{2\Re{z}}\right) = S_0^{2\Re{z}} e^{s\kappa(2\Re{z})} \leq S_0^{2R} \left(e^{T\kappa(2R)}\vee 1\right) \]
and Fubini's Theorem, we obtain from~\eqref{e:PboundsE1} that
\begin{equation}\label{eq:PboundsE2}
\begin{aligned}
E\left(\covar{l(z)}{\overline{l(z)}}_T\right) & \leq \left(S_0^{2R}\left(e^{T\kappa(2R)}\vee 1\right)\right)\left(\int_0^T \abs{\alpha(z,s)}^2 \abs{\kappa(2\Re{z})-2\Re{\kappa(z)}} \,ds \right.\\
& \qquad +\int_0^T \abs{g(z,s)}^2 \abs{\kappa(2)-2\kappa(1)}\,ds\\
& \qquad + \left. \int_0^T 2 \abs{\alpha(z,s)} \abs{g(z,s)} \abs{\kappa(z+1)-\kappa(z)-\kappa(1)} \,ds \right).
\end{aligned}
\end{equation}
To prove Assertion 3, it therefore suffices to show that all integrals in \eqref{eq:PboundsE2} are $p$-integrable. For the first one, we have
\begin{multline}\label{eq:Pbounds3}
 \int_0^T \abs{\alpha(z,s)}^2 \abs{\kappa(2\Re{z})-2\Re{\kappa(z)}} \,ds \\ \leq \abs{\kappa(2R)}\int_0^T \abs{\alpha(z,s)}^2 \,ds + 2\int_0^T \abs{\alpha(z,s)}^2 \abs{\Re{\kappa(z)}}\, ds
\end{multline}
Let us first consider the second integral in~\eqref{eq:Pbounds3}. Inserting the representation~\eqref{sol_Alpha} for $\alpha$, we obtain
\begin{align*}
 & \int_0^T \abs{\Re{\kappa(z)}} \abs{\alpha(z,s)}^2 \,ds  \\
 & \leq \int_0^T \abs{\Re{\kappa(z)}} e^{2\Re{\kappa(z)}(T-s)} \,ds \\
 & \quad +2\int_0^T \abs{\Re{\kappa(z)}} e^{\Re{\kappa(z)}(T-s)} \abs{\kappa(1)} \left(\int_s^T e^{\Re{\kappa(z)}(\tau-s)} \abs{g(z,\tau)} \,d\tau\right) \,ds \\
 & \quad +\int_0^T \abs{\Re{\kappa(z)}} \abs{\kappa(1)}^2 \left(\int_s^T e^{\Re{\kappa(z)}(\tau-s)} \abs{g(z,\tau)} \,d\tau \right)^2 \,ds \\
 & \leq \int_0^T \abs{\Re{z}} e^{2\Re{\kappa(z)}(T-s)}\,ds \\
 & \quad +2 \int_0^T \abs{\Re{\kappa(z)}} e^{\Re{\kappa(z)}(T-s)} \abs{\kappa(1)} \left(\int_s^T e^{2\Re{\kappa(z)}(\tau-s)} \,d\tau\right)^{\frac{1}{2}} \\
 & \qquad\quad \times \left(\int_s^T \abs{g(z,\tau)}^2 \,d\tau\right)^{\frac{1}{2}} \,ds \\
 & \quad+ \int_0^T \abs{\Re{\kappa(z)}} \abs{\kappa(1)}^2 \left(\int_s^T e^{2\Re{\kappa(z)}(\tau-s)} \,d\tau\right) \left(\int_s^T \abs{g(z,\tau)}^2 \,d\tau\right) \,ds,
\end{align*}
where we applied H\"older's inequality twice in the last step. Using that $\Re{\kappa(z)}$, $z\in\Sf$, is bounded from above by the constant~$b_1\geq 0$, it is easily seen by elementary integration that the integrals of the form \[ \int_a^T \abs{\Re{\kappa(z)}} e^{m\Re{\kappa(z)}s}\,ds, \quad 0\leq a\leq T, \quad m\in\{1,2\},\] are uniformly bounded on $\Sf$. Moreover, the terms $\int_s^T \abs{g(z,\tau)}^2\,d\tau$ are bounded by $\int_0^T \abs{g(z,\tau)}^2\,d\tau$, which is $p$-integrable since $\varphi$ is a \dstr. The $p$-integrability of $\left(\int_0^T \abs{g(z,\tau)}^2\,d\tau\right)^{\frac{1}{2}}$ follows after an application of H\"older's inequality. Altogether, this yields that the second integral in~\eqref{eq:Pbounds3} is $p$-integrable. The $p$-integrability of the first integral in~\eqref{eq:Pbounds3} follows analogously by inserting the representation for~$\alpha$ from \eqref{sol_Alpha} and exploiting that $\varphi$ is a \dstr. By the latter fact, we obtain directly that the second integral in~\eqref{eq:PboundsE2} is $p$-integrable as well. To deal with the third one, we use the inequality \[\abs{\kappa(z+1)-\kappa(z)-\kappa(1)}^2 \leq \left(\kappa(2\Re{z})-2\Re{\kappa(z)}\right) \left(\kappa(2)-2\kappa(1)\right)\] established in \cite[Lemma~3.4]{hub06} and apply H\"older's inequality to conclude that
\begin{equation*}
\begin{aligned}
& \int_0^T \abs{\alpha(z,s)} \abs{g(z,s)} \abs{\kappa(z+1)-\kappa(z)-\kappa(1)} \,ds \\
& \leq \left(\int_0^T \abs{g(z,s)}^2\,ds\right)^{\frac{1}{2}}\\
& \quad \times \left( \int_0^T \abs{\alpha(z,s)}^2 \abs{\kappa(2\Re{z})-2\Re{\kappa(z)}} \abs{\kappa(2)-2\kappa(1)} \,ds \right)^{\frac{1}{2}}.
\end{aligned}
\end{equation*}
Using H\"older's inequality, this proves Assertion 3, because the squares of both integrals on the right-hand side have already been shown to be $p$-integrable.
\item Note that in~\eqref{e:PboundsE1} the only stochastic terms are given by $S_{-}^{2\Re{z}}=S_{-}^{2R}$. Since this process is locally bounded, almost all of its paths are bounded on $[0,T]$. The estimates from the proof of Assertion~\ref{l:boundEx} therefore also show that Assertion~\ref{e:boundCoverOmega} holds. \qedhere
\end{enumerate}
\end{prf}

The estimates in Lemma \ref{l:bounds} immediately yield the following

\begin{cor}\label{c:lSqIntegr}
For all $z\in\Sf$, the process $l(z)$ defined in~(\ref{defn_zErrorProc}) is a square-integrable martingale.
\end{cor}

\begin{prf}
Recall that $l(z)$ has already shown to be a local martingale in Theorem~\ref{thmCorrectionErrorFixedZ}. Since $[l(z),\overline{l(z)}]-\langle l(z),\overline{l(z)}\rangle$ is a local martingale as well, the assertion follows from  \cite[I.4.50(c)]{js03} by localization, monotone convergence and the fact that the right-hand side of~\eqref{eq:PboundsE2} is finite for all $z\in R+i\R$.\end{prf}

The next two propositions show that the candidates proposed in Equations~\eqref{eq:CandL} and~\eqref{eq:CandCovarL} indeed coincide with the desired martingale and its quadratic variation of Ansatz~\eqref{eq:AnsatzBigL}.

\begin{prop}\label{p:L}
 The process~$L$ defined by
 \begin{equation}\label{d:ErrorProc}
     L_t := \int_{\rmif}^{\rpif} l(z)_t\, p(z)\, dz - c
 \end{equation}
  is a real-valued square-integrable martingale with $L_T = H - c - \varphi\sint S_T.$
\end{prop}

\begin{prf}
 First note that by Lemma~\ref{l:bounds}(\ref{e:bndAlpha}) and Proposition~\ref{p:fubini}, the integral in~(\ref{d:ErrorProc}) is well-defined. Fubini's Theorem and dominated convergence show that $\int_{\rmif}^{\rpif} S_t^z\alpha(z, t)\,p(z)\,dz$ is an adapted \cadlag{} process. For $t\in[0,T]$, H\"older's inequality and another application of Fubini's Theorem yield
 \begin{align*}
     E\left(\left| \int_\rmif^\rpif l(z)_t\,p(z)\,dz \right|^2\right) & \leq  E\left(\int_\rmif^\rpif \left|l(z)_t\right|^2\,\left|p(z)\right|\,|dz|\right) \left(\int_\rmif^\rpif \left|p(z)\right|\,|dz|\right) \\
     & = \left(\int_\rmif^\rpif \ew{\left|l(z)_t\right|^2}\,\left|p(z)\right|\,|dz|\right) \left(\int_\rmif^\rpif \left|p(z)\right|\,|dz|\right).
 \end{align*}
 Since $l(z)$ is a square-integrable martingale by Corollary~\ref{c:lSqIntegr}, \cite[I.4.2 and~I.4.50(b)]{js03} imply
 \begin{equation} \label{eq:boundExpL}
     E\left(\left|l(z)_t\right|^2\right) = \left|l(z)_0\right|^2 + E\left(\covar{l(z)}{\overline{l(z)}}_t\right) \leq S_0^{2R} \abs{\alpha(z,0)}^2 + E\left(\covar{l(z)}{\overline{l(z)}}_T\right).
 \end{equation}
 Because the right-hand side of~\eqref{eq:boundExpL} is $p$-integrable by Lemma~\ref{l:bounds}(\ref{e:bndAlpha},\ref{l:boundEx}) and independent of $t\in[0,T]$, we conclude that $L$ is indeed square-integrable, i.e.,
 \begin{equation*}
     \sup_{t\in[0,T]} E\left(\left|L_t\right|^2\right) < \infty.
 \end{equation*}
 In order to show the martingale property of~$L$, consider arbitrary $0\leq s\leq t\leq T$ and $F\in\F_s$. By Fubini's Theorem and the martingale property of $l(z)$, we have
 \begin{align*}
     E\left((L_t-L_s)1_F\right) &= E\left(\int_\rmif^\rpif \left(l(z)_t - l(z)_s\right)1_F \,p(z)\,dz\right) \\
     &= \int_\rmif^\rpif E\left(\left(l(z)_t-l(z)_s\right)1_F\right) \,p(z)\,dz = 0,
 \end{align*}
 and hence $E(L_t|\F_s) = L_s$. Since $\alpha(z,T)=1$, it follows from Assumption~\ref{a:intReprF} and Proposition~\ref{p:fubini} that $L_T$ is given by the asserted, real-valued random variable. The martingale property of $L$ then yields that $L_t$ is real-valued for all $t \in [0,T]$, which completes the proof.
\end{prf}

\begin{prop}\label{p:CovarL}
 The predictable quadratic variation $\covar{L}{L}$ of the process~$L$ defined in~(\ref{d:ErrorProc}) is given by
 \begin{equation} \label{eq:predCovarBigL}
     \covar{L}{L}_t = \int_{\rmif}^{\rpif} \int_{\rmif}^{\rpif} \covar{l(y)}{l(z)}_t \, p(y)p(z)\,dydz.
 \end{equation}
\end{prop}

\begin{prf}
By Proposition \ref{p:L} and \cite[I.4.2 and~I.4.50(b)]{js03}, it suffices to prove that the candidate $C := \int_\rmif^\rpif \int_\rmif^\rpif \covar{l(y)}{l(z)} \, p(y)p(z)\,dydz$ is a predictable process of finite variation such that $L^2-C$ is a local martingale. First, note that the integral in~\eqref{eq:predCovarBigL} is well-defined by Lemma~\ref{l:bounds}\eqref{e:boundCoverOmega}. Moreover, since $t\mapsto\covar{l(y)}{l(z)}_t$ is continuous by Proposition~\ref{p:predCovarL} and hence predictable for all $y,z\in\Sf$, the process $C$ is predictable as well by Fubini's Theorem. To see that $C$ is of finite variation, note that it is a linear combination of expressions of the form \[ \int_\rmif^\rpif \int_\rmif^\rpif \Re{\covar{l(y)}{l(z)}} \,h(y)j(z)\,dy dz \] or \[\int_\rmif^\rpif \int_\rmif^\rpif \Im{\covar{l(y)}{l(z)}} \,h(y)j(z)\,dy dz \] for $h,j\in \left\{\Re{p}^+,\Re{p}^-,\Im{p}^+,\Im{p}^-\right\}$. In view of~\eqref{eq:decompdCovarL}, we obtain that
 \begin{align*}
     & \int_\rmif^\rpif \int_\rmif^\rpif \Re{\covar{l(y)}{l(z)}} \,h(y)j(z)\,dy dz \\
     & = \frac{1}{2} \int_\rmif^\rpif\int_\rmif^\rpif \covar{l(y)+\overline{l(z)}}{\overline{l(y)+\overline{l(z)}}} \,h(y)j(z)\,dydz \\
     & \quad -\frac{1}{2}\left( \int_\rmif^\rpif \int_\rmif^\rpif \left(\covar{l(y)}{\overline{l(y)}}+\covar{l(z)}{\overline{l(z)}}\right) \,h(y)j(z)\,dy dz \right),
 \end{align*}
 which is the difference of two increasing adapted processes and hence of finite variation. The argument applies analogously to $\Im{\covar{l(y)}{l(z)}}$. To show the martingale property of~$L^2-C$, we can assume without loss of generality that $c=0$. Observe that
 \begin{equation*}%\label{eq:ExpLMinusC}
 \begin{aligned}
 E\left(\abs{L_t^2-C_t}\right) &= E\left(\abs{\int_\rmif^\rpif \int_\rmif^\rpif \big(l(y)_tl(z)_t - \covar{l(y)}{l(z)}_t\big) \,p(y)p(z)\,dydz} \right) \\
 & \leq \int_\rmif^\rpif \int_\rmif^\rpif \big((E\left(\abs{l(y)_tl(z)_t}\right) + E\left(\abs{\covar{l(y)}{l(z)}_t}\right)\big)\abs{p(y)}\abs{p(z)}|dy||dz|.
 \end{aligned}
 \end{equation*}
 Moreover, $E(\abs{l(y)_t l(z)_t}) \leq E(\abs{l(y)_t}^2)^{1/2} E(\abs{l(z)_t}^2)^{1/2}$ and $E(\abs{\covar{l(y)}{l(z)}_t})$ are twice $p$-integrable by Lemma~\ref{l:bounds}\eqref{l:boundEx}, H\"older's inequality and the fact that the right-hand side of~\eqref{eq:boundExpL} is $p$-integrable. This shows that $L^2_t-C_t \in L^1(P)$. For $0\leq s\leq t\leq T$ and $F\in\F_s$, we can therefore apply Fubini's Theorem to obtain
 \begin{align*}
     E\left(\left(L^2_t-C_t\right)1_F\right) &= \int_\rmif^\rpif \int_\rmif^\rpif E\left(\left(l(y)_t l(z)_t - \covar{l(y)}{l(z)}_t\right)1_F\right) \,p(y)p(z)\,dydz \\
     &= \int_\rmif^\rpif \int_\rmif^\rpif E\left(\left(l(y)_s l(z)_s - \covar{l(y)}{l(z)}_s\right)1_F\right) \,p(y)p(z)\,dydz \\
     &= E\left(\left(L^2_s-C_s\right)1_F\right),
 \end{align*}
 where we used in the second step that $l(y)l(z)-\covar{l(y)}{l(z)}$ is a martingale by Corollary~\ref{c:lSqIntegr} and \cite[I.4.2 and~I.4.50(b)]{js03}. Hence, $E\left(\left.L^2_t-C_t\right|\F_s\right) = L^2_s - C_s$, which completes the proof.
\end{prf}

Now we can finally prove our main Theorem~\ref{thm:MainResult} by combining the preceding results.

\begin{prf}[Proof of Theorem~\ref{thm:MainResult}]
In view of Lemma \ref{l:L(S)}, the mean squared hedging error corresponding to the endowment/strategy pair $(c,\varphi)$ is well-defined. By Assumption~\ref{a:intReprF},  Proposition~\ref{p:fubini}, the definition of $l(z)$ in~\eqref{defn_zErrorProc} and Proposition \ref{p:L}, we have
$$E((H-c-\varphi \sint S_T)^2) = E(L_T^2) =E(L_0^2) + E(\covar{L}{L}_T),$$
where the second equality follows from \cite[I.4.2 and~I.4.50(b)]{js03}. Now notice that by definition,
$$L_0^2 = \left(\int_\rmif^\rpif S_0^z \alpha(z,0)\,p(z)\,dz - c\right)^2.$$
In view of Lemma \ref{l:bounds}(3), Fubini's Theorem and Proposition~\ref{p:CovarL} yield
$$ E(\covar{L}{L}_T) = \int_\rmif^\rpif \int_\rmif^\rpif E(\covar{l(y)}{l(z)}_T)\,p(y) p(z)\,dy dz.$$
Since $E\left(S_{t-}^{y+z}\right) = S_0^{y+z} e^{t\kappa(y+z)}$, and because the continuous functions $t \mapsto |g(z,t)|$, $t \mapsto |\alpha(z,t)|$ and
\[ t \mapsto E\left(\left|S_{t-}^{y+z}\right|\right) \leq S_0^{2R} e^{t\kappa(2R)}\]
are bounded on $[0,T]$, Proposition~\ref{p:predCovarL} and another application of Fubini's Theorem complete the proof.
\end{prf}
\end{appendix}

\section*{Acknowledgments}
We thank three anonymous referees for their constructive comments, which led, in particular, to the numerical example in Section~\ref{ss:CGMYE}. Moreover, the third author gratefully acknowledges partial financial support through \emph{Sachbeihilfe \mbox{KA 1682/2-1}} of the \emph{Deutsche Forschungsgemeinschaft}. The fourth author gratefully acknowledges financial support by the \textit{National Centre of Competence in Research ``Financial Valuation and Risk Management''} (NCCR FINRISK), Project D1 (Mathematical Methods in Financial Risk Management). The NCCR FINRISK is a research instrument of the \textit{Swiss National Science Foundation}.

\addcontentsline{toc}{section}{References}
\bibliography{references}

\end{document}